\documentclass[prb,aps,twocolumn,showpacs]{revtex4}

\usepackage{graphicx}

\begin{document}

\title{Effects of inter-site Coulomb interactions on ferromagnetism: \\
       Application to Fe, Co and Ni}

\author{     Cyrille Barreteau and Marie-Catherine Desjonqu\`eres }
\affiliation{Commissariat \`a l'Energie Atomique, DSM/DRECAM/SPCSI, \\
             Centre d'Etudes de Saclay, F-91191 Gif sur Yvette, France }

\author{     Andrzej M. Ole\'{s} }
\affiliation{Marian Smoluchowski Institute of Physics, Jagellonian
             University, Reymonta 4, PL-30059 Krak\'{o}w, Poland}

\author{     Daniel Spanjaard }
\affiliation{Laboratoire de Physique des Solides,
             Universit\'e Paris Sud, Batiment 510, F-91405 Orsay, France}

\date {\today}

\begin{abstract}
We reanalyze the condition for metallic ferromagnetism in the
framework of the tight-binding approximation and investigate the
consequences of the inter-site Coulomb interactions using the
Hartree-Fock approximation. We first consider a non-degenerate $s$
band and we show that the inter-site interactions modify the
occurrence of ferromagnetism, and we derive a generalized Stoner
criterion. We analyze the main effects due to the renormalization
of the hopping integrals by the inter-site Coulomb interactions.
These effects are strongly dependent on the relative values of the
inter-site electron-electron interactions and on the shape of the
density of states as illustrated by a study of cubic crystals from
which we establish general trends. We then investigate a realistic
$spd$ tight-binding model, including intra (Coulomb and exchange)
and inter-site charge-charge Coulomb integrals. This model is used
to study the electronic structure (band structure, densities of
states, magnetic moment) of bulk ferromagnetic $3d$ transition
metals Fe(bcc), Co(hcp and fcc) and Ni(fcc). An excellent
agreement with local spin density functional calculations is
obtained for the three metals, in particular concerning the
relative widths of the majority and minority spin bands. Thus our
tight-binding Hartree-Fock model provides a consistent
interpretation of this effect.
\end{abstract}

\maketitle


\section{Introduction}

The origin of ferromagnetism in itinerant systems remains one of
the open questions in the condensed matter theory. Even though
spin-density functional theory gives correctly several predictions
concerning the stability of ferromagnetism,\cite{Mor78} it is
interesting to develop simple models which point out the important
physical parameters governing the ferromagnetic instabilities.
Ferromagnetism may occur in two ways: either the paramagnetic (PM)
state gets unstable against the ferromagnetic (FM) state for
particular values of parameters (so-called Stoner instability), or
the strongly polarized FM state, in most cases saturated ferromagnetic
(SF) state, has the lowest energy beyond some values of
electron-electron interactions, but the PM state is still locally stable
in a range of parameters.

It is certainly instructive to understand first the possible
mechanisms of ferromagnetism in the case of a narrow $s$ band
studied in the tight-binding model using the Hartree-Fock
approximation (HFA). When only the on-site matrix element U of the
Coulomb interaction are taken into account the Stoner instability
occurs when $U$ satisfies the well known Stoner criterion
$UN(E_F)>1$, where $N(E_F)$ is the density of states at the Fermi
level per spin, and the majority and minority spin bands are
rigidly shifted relative to each other. The influence of the
two-site matrix elements of the Coulomb interaction has been
thoroughly studied in the pioneering work by Hirsch {\it et al.}
already over a decade ago.\cite{Hir89,Hir90} These ideas were
further developed and qualitatively new effects were found, both
within the HFA, and by going beyond
it.\cite{Ole92,Str94,Hir96,Hir99,Pou00} In particular, Hirsch {\it
et al.}\cite{Hir89,Hir90,Hir96,Hir99} have pointed out that the
renormalization of the hopping integrals resulting from the
Hartree-Fock decoupling of the two-body inter-site term of the
hamiltonian plays a role in the occurrence of ferromagnetism by
changing the bandwidths of majority and minority spin bands in a
different way, thus modifying the Stoner condition. However,
Hirsch {\it et al.}\cite{Hir89,Hir90,Hir96,Hir99} have mainly
emphasized the effect of exchange integrals and have only
considered a constant density of states or that of a linear chain.
Indeed, the non-degenerate model with exchange interactions was
proposed to provide explanation of certain itinerant systems, such
as EuB$_6$.\cite{Hireb} Obviously, the conclusions that can be
drawn from such a model depend critically on the numerical values
of the parameters and we expect the classical charge-charge
inter-site Coulomb interaction $V$ to be larger than the exchange
one.\cite{Str94} Consequently, we have found interesting to
revisit this model by considering more realistic relative values
of the Coulomb matrix elements first in the case of a constant
density of states, then for the densities of states of real cubic
lattices. We will see that in this last case the physics of FM
instabilities is significantly modified. Therefore a quantitative
study of the influence of inter-site Coulomb interactions in FM
transition metals needs an accurate description of the density of
states and realistic interaction parameters.

However ferromagnetism is found usually in systems with degenerate
orbitals and since the early work of Hubbard\cite{Hub63} a variety
of theoretical attempts have been undertaken to understand to what
extent the orbital degeneracy might play an essential role in
particular in the ferromagnetism of Fe, Co and Ni. The dominating
point of view at present is that the degeneracy of $d$ orbitals is
crucial,\cite{Mat81,Ole84,Sto90} since local moments can form
\cite{Stollhoff81} and survive above the Curie temperature due to
intra-atomic exchange integrals. Here we will investigate the
effect of on-site and inter-site interactions in degenerate $spd$
bands, considering realistic parameters. It is clear that the most
important matrix elements of the Coulomb interaction involved in
FM instabilities contain four $d$ orbitals centered on at most
two neighboring sites. The relative values of the matrix elements
can be inferred from the results of explicit calculations using
atomic wave functions. From these calculations\cite{Cormier} it
turns out that the largest on-site matrix elements are those
introduced already in our previous paper,\cite{Barreteau00} while
the only non negligible inter-site Coulomb interactions are of
electrostatic (i.e., charge-charge) type, arising from the
interaction between two electrons in orbitals centered at two
neighboring sites. As for an $s$ band, the Hartree-Fock decoupling
of these latter terms renormalizes the hopping integrals and we
will show that they are responsible for the different $d$
bandwidths for up and down spins which are obtained in local spin
density functional calculations for the $3d$ FM elements (Fe, Co,
Ni). Furthermore, the nearest neighbor distances being very close
for the three metals, it is expected that the nearest neighbor
Coulomb interactions should not vary significantly from Fe to Co
and Ni. It will be seen that, under this assumption, an excellent
agreement is found between local spin density and our
tight-binding Hartree-Fock (TBHF) calculations for the band
structure, the densities of states, as well as the magnetic moment
of the three elements.

The paper is organized as follows. We recall (Sec. II) the $s$
band (extended Hubbard Hartree-Fock) model in which all
interactions are limited to first nearest neighbors and derive an
analytic generalized Stoner criterion. including electrostatic as
well as exchange interactions. Then we revisit the simple model of
a constant density of states which can be solved
analytically,\cite{Hir89,Hir96,Hir99} and we show that the onset
of the Stoner instability and the condition of occurrence of
saturated ferromagnetism are strongly dependent on the ratio of
the two-site electrostatic interaction and exchange interaction.
Finally, in order to illustrate the role of the shape of the
density of states, we consider the case of three-dimensional
lattices: simple (sc), body centered (bcc) and face centered (fcc)
cubic lattice for reasonable parameters, with particular emphasis
on the role played by inter-site Coulomb interactions. In Sec.
III, the multi-band $spd$ TBHF model, already described by
Barreteau {\it et al.},\cite{Barreteau00} is extended by including
the electrostatic inter-site Coulomb matrix elements, and the
determination of the parameters is discussed. This model is
finally used in Sec. IV to analyze the band structures of FM
transition metals: Fe, Co and Ni. We will show that the main
effects which can be understood for the $s$ band using some
analytic arguments apply also qualitatively to $3d$ transition
metals. The paper is concluded in Sec. V.

\section{Ferromagnetism in a non-degenerate $s$ band }

\subsection{The tight-binding Hartree-Fock model}

We start from a tight-binding model for an $s$ band and assume
that the set of atomic ($s$) orbitals $\phi_i(r)$ centered at each
site $i$ is orthogonal. We consider the same Hamiltonian as
Hirsch\cite{Hir99} with inter-site interactions limited to first
nearest neighbors which, in the second quantization formalism, can
be written:
\begin{eqnarray}
H_s\!&=&\!-t\sum_{i,j\neq i,\sigma}a^{\dagger}_{i\sigma}a^{}_{j\sigma}
  +\frac{U}{2}\sum_{i,\sigma}n^{}_{i\sigma}n^{}_{i-\sigma} \nonumber \\
  &+&\frac{V}{2}\sum_{i,j\neq i,\sigma,\sigma'}a^{\dagger}_{i\sigma}
  a^{\dagger}_{j\sigma'}a^{}_{j\sigma'}a^{}_{i\sigma}
                                           \nonumber \\
 &+&\!\frac{J}{2}\!\!\sum_{i,j\neq i,\sigma,\sigma'}\!
    a^{\dagger}_{i\sigma}a^{\dagger}_{j\sigma'}a^{}_{i\sigma'}a^{}_{j\sigma}
    \nonumber  \\
 &+&\frac{J'}{2}\sum_{i,j\neq i,\sigma}a^{\dagger}_{i\sigma}
       a^{\dagger}_{i-\sigma}a^{}_{j-\sigma}a^{}_{j\sigma},
\label{hubs}
\end{eqnarray}
where $a^{\dagger}_{i\sigma}$ is the creation operator of an
electron with spin $\sigma$ in the atomic orbital centered at site
$i$, $n_{i\sigma}=a^{\dagger}_{i\sigma}a^{}_{i\sigma}$, and $-t$
is the hopping integral between nearest neighbors. The Coulomb
interactions are described by the leading on-site term $\propto
U$, and by the two-site terms: charge-charge interactions $\propto
V$, exchange interactions $\propto J$, and the 'pair hopping' term
$\propto J'$,
\begin{eqnarray}
\label{us} U &=& \langle\phi_{i}(r) \phi_{i}(r') | \frac{1}{|{\bf
r-r'}|} |
             \phi_{i}(r) \phi_{i}(r') \rangle ,   \\
\label{vs} V &=& \langle\phi_{i}(r) \phi_{j}(r') |\frac{1}{|{\bf
r-r'}|} |
             \phi_{i}(r) \phi_{j}(r') \rangle,      \\
\label{js} J &=& \langle\phi_{i}(r) \phi_{j}(r') |\frac{1}{|{\bf
r-r'}|} |
             \phi_{j}(r) \phi_{i}(r') \rangle,      \\
\label{ds} J'&=& \langle\phi_{i}(r) \phi_{i}(r') | \frac{1}{|{\bf
r-r'}|} |
             \phi_{j}(r)\phi_{j}(r')  \rangle.
\end{eqnarray}
We will consider in most cases the realistic relation $J=J'$, as
obtained for real wave functions from Eqs. (\ref{js}) and (\ref{ds}).

In the HFA the two-body terms are decoupled in the following way:
\begin{eqnarray}
a^{\dagger}_{\alpha}a^{\dagger}_{\beta}a^{}_{\gamma}a^{}_{\delta}&=&
\langle a^{\dagger}_{\alpha}a^{}_{\delta}\rangle
        a^{\dagger}_{\beta}a^{}_{\gamma}
+\langle a^{\dagger}_{\beta}a^{}_{\gamma}\rangle         \nonumber \\
&-&\langle a^{\dagger}_{\alpha}a^{}_{\gamma}\rangle
        a^{\dagger}_{\beta}a^{}_{\delta}
-\langle a^{\dagger}_{\beta}a^{}_{\delta}\rangle
        a^{\dagger}_{\alpha}a^{}_{\gamma}                \nonumber \\
&-&\langle a^{\dagger}_{\alpha}a^{}_{\delta}\rangle \langle
           a^{\dagger}_{\beta}a^{}_{\gamma}\rangle
 + \langle a^{\dagger}_{\alpha}a^{}_{\gamma}\rangle \langle
           a^{\dagger}_{\beta}a^{}_{\delta}\rangle,
\label{hfa}
\end{eqnarray}
where the indices denote an atomic spin-orbital. Note also that
the spin conservation implies that any average
$\langle a^{\dagger}_{\alpha}a^{}_{\beta}\rangle$ vanishes when $\alpha$
and $\beta$ have different spins.

Consequently the HFA leads to a one-particle Hamiltonian:
\begin{equation}
H_s^{\rm HF}=-\sum_{i,j\neq i,\sigma}t_{\sigma}
   a^{\dagger}_{i\sigma}a^{}_{j\sigma}
  +\sum_{i\sigma}\varepsilon_{\sigma}n^{}_{i\sigma}-E_{dc},
\label{hfs}
\end{equation}
with the following spin-dependent hopping integrals and orbital energies,
\begin{eqnarray}
\label{ts}
t_{\sigma}&=&t+(V-J)I_{\sigma}-(J+J')I_{-\sigma},         \\
\label{es}
\varepsilon_{\sigma}&=&z(V-J)n+(U+zJ)n_{-\sigma},
\end{eqnarray}
$E_{dc}$ stands for the double counting energy terms and $z$ is the
number of nearest neighbors. Here we assume that the system consists
of equivalent atoms, then the occupation numbers:
$n_{\sigma}=\langle n_{i\sigma}\rangle$ for electrons of spin
$\sigma$ and the total band filling $n=\sum_{\sigma}n_{\sigma}$
are the same at each site $i$. In this model $I_{\sigma}=\langle
a^{\dagger}_{i\sigma}a^{}_{j\sigma}\rangle$ do not depend on the
bond and can be easily obtained from the density of states (per atom)
$N_{\sigma}(E)$. Indeed, choosing the origin of energies at the center
of gravity of $N_{\sigma}(E)$, it follows from Eq. (\ref{hfs}) that:
\begin{equation}
\sum_{n occ} E_{n\sigma}=\int^{E_{F\sigma}}_{-\infty}
EN_{\sigma}(E)\;dE=-zt_{\sigma}I_{\sigma},
\label{Isigma}
\end{equation}
where $E_{n\sigma}$ are the eigenenergies for spin
$\sigma$. Note that $I_{\sigma}=I(n_{\sigma})$ since the
renormalization of the hopping integrals leads to a simple energy
rescaling of the density of states without changing its shape.

\subsection{Generalized Stoner criterion and condition for
saturated ferromagnetism}

The magnetic energy as a function of the magnetic moment
$m=n_{\uparrow}-n_{\downarrow}$:
\begin{equation}
E_{mag}(m)=\langle H^{\rm HF}_s(m)\rangle-\langle H^{\rm HF}_s(0)\rangle,
\end{equation}
where $\langle H^{\rm HF}_s(m)\rangle$ is the Hartree-Fock energy of a
state with magnetic moment $m$, is easily expressed in terms of the
function $I$. Let us introduce the following notations:
\begin{equation}
I_{\sigma}= I(n_{\sigma})=I\Big(\frac{n+\sigma m}{2}\Big),
\end{equation}
with $\sigma=+1(-1)$ for up(down) spin. Then the magnetic energy per
atom is given by:
\begin{eqnarray}
E_{mag}(m)&=&2zt\Big[I_0-\frac{1}{2}(I_{\uparrow}+I_{\downarrow})\Big]
           -\frac{1}{4}(U+zJ)m^2                              \nonumber \\
&-&\frac{1}{2}z(V-J)(I_{\uparrow}^2+I_{\downarrow}^2-2I_0^2)  \nonumber \\
&+&z(J+J')(I_{\uparrow}I_{\downarrow}-I_0^2), \label{emagg}
\end{eqnarray}
where $I_0$ refers to the PM state, i.e., $I_0=I(n/2)$.

Let us first derive the condition under which the PM state becomes
unstable (Stoner instability). This instability is obtained from
the Taylor expansion of $I_{\sigma}$ for small magnetization $m$.
Substituting this expansion for $I_{\uparrow}$ and $I_{\downarrow}$
into (\ref{emagg}) one finds to second order in $m$:
\begin{eqnarray}
E_{mag}(m)&=&[-zt{I''}_0-z(V-J)({I'}_0^2+I_0{I''}_0)  \nonumber \\
&+&z(J+J')(I_0{I''}_0-{I'}_0^2)  \nonumber \\
&-&(U+zJ)]\frac{m^2}{4},
\label{eq:emag2}
\end{eqnarray}
where $I'_0$ and $I''_0$ are the first and second derivatives of
$I_{\sigma}$ at $n_{\sigma}=n/2$. The Stoner instability occurs
when the coefficient of $m^2$ is negative, i.e.,
\begin{equation}
zI_0''[t+(V-2J-J')I_0]+z(V+J')I_0'^2+U+zJ > 0.
\label{stoner}
\end{equation}

>From Eq. (\ref{ts}) it is seen that the term between brackets is
the hopping integral $t_{\rm PM}$ in the PM state. Using now
Eq. (\ref{Isigma}) it can be shown, with obvious notations, that:
\begin{eqnarray}
I'_0 &=&-\frac{E_F}{zt}=-\frac{E_F^{\rm PM}}{zt_{\rm PM}},  \nonumber \\
I''_0 &=&-\frac{1}{ztN(E_F)}=-\frac{1}{zt_{\rm PM}N_{\rm
PM}(E_F^{\rm PM})},
\label{IPIPP}
\end{eqnarray}
where $N_{\rm PM}(E_F^{\rm PM})$ is the density of states per spin in the PM state
at the Fermi level $E_F^{\rm PM}$.
Note that in these equations the energies must be referred to the
center of gravity of the band. The inequality (\ref{stoner}) can be
now rewritten as a {\it generalized Stoner criterion}:
\begin{equation}
U_{eff}N_{\rm PM}(E_F^{\rm PM}) > 1,
\end{equation}
with:
\begin{equation}
U_{eff}=U+zJ+z(V+J')\Big(\frac{E_F^{\rm PM}}{zt_{\rm PM}}\Big)^2.
\end{equation}

This generalizes the criterion derived by Hirsch (Eq. (22)) in
Ref. \onlinecite{Hir99}] for the particular case of a constant
density of states. Accordingly the influence of the inter-site
exchange integral $J$ is to act in favor of the FM state for any
band filling since it increases $U_{eff}$ and decreases the
bandwidth of the PM state [see Eq. (\ref{ts})]. Let us now examine
the effect of $V$ and $J'$. At low and high band fillings the
renormalization of the hopping integral in the PM state tends to
zero since $I_0$ vanishes. As a consequence, due to the term
proportional to $V+J'$ in $U_{eff}$ the PM state is more easily
destabilized for low values of $n$ since, in this case, the ratio
$(E_F^{\rm PM}/zt_{\rm PM})^2$ is close to unity, the bottom of
the band being at $E=-zt_{\rm PM}$. This is also true when $n$
approaches $n=2$ for simple and body centered cubic lattices and
this tendency is weakened for the face centered cubic lattice
since $(E_F^{\rm PM}/zt_{\rm PM})^2=1/9$. Around half filling
$E_F^{\rm PM}$ is small so that $U_{eff}\simeq U+zJ$ and for
realistic values of $V$ (i.e., $V>2J+J'$) the PM band is
broadened. Moreover its width, like $I_0$, is there at its maximum
and increases with $V$. Consequently, the PM state is less easily
destabilized around half filling.

Finally it is also interesting to derive the value of $U$ above
which the SF state with moment $m_s$ ($m_s=n$ when $n\leq 1$,
$m_s=2-n$ when $n\geq 1$) becomes the most stable. This is done
simply by looking for the minimum value of $U$ for which the
function $E_{mag}(m)$ (Eq.(\ref{emagg})) takes its minimum value
at $m_s$ in the domain $[0,m_s]$.

\subsection{The magnetic instabilities for a constant density of states
            revisited}

In order to obtain a physical insight into the mechanism of FM
instability in the $s$ band model (Eq. (\ref{hfs})), we discuss
first the case of a constant density of states (per spin, with the
zero of energy at $\varepsilon_{\sigma}$),
\begin{equation}
N_{\sigma}(E)=\frac{1}{W_{\sigma}}, \hskip .7cm {\rm for}
                                    \hskip .3cm |E|<W_{\sigma}/2,
\label{const}
\end{equation}
with $W_{\sigma}=2zt_{\sigma}$ (in the following we assume $z=6$).
Using Eq. (\ref{Isigma}) it is found that $I_{\sigma}$ has a very
simple analytical expression:
\begin{equation}
I_{\sigma}=n_{\sigma}(1-n_{\sigma}).
\label{fock}
\end{equation}
Let us first discuss the renormalization of the hopping integrals
for a more than half-filled band, as in the late $3d$ transition
metals, in the PM phase as well as in the SF phase which becomes
stable when $U$ is large enough. It is clear that the majority
spin up band is narrower than the minority spin down one and it
can be shown easily that the hopping integral found in the PM
phase lies always in between the ones for the majority and
minority spins in the SF phase at least in the realistic case
\cite{Str94} where $V > J$. We will see in Sec. III that this
holds also qualitatively for Fe, Co and Ni.

\begin{figure}
\includegraphics*[width=8.5cm]{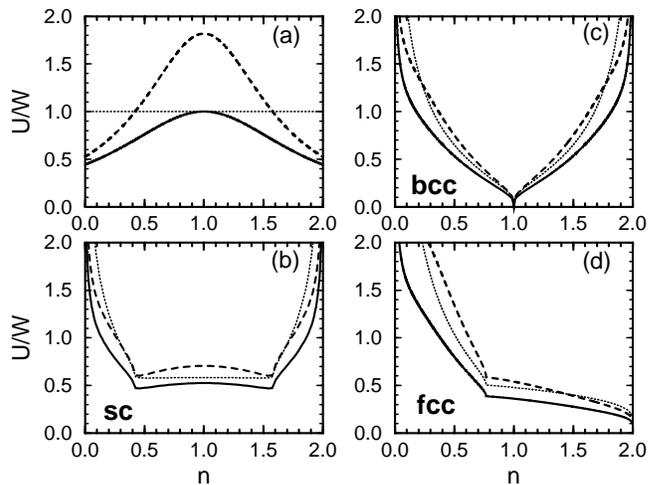}
\caption {Critical value of $U/W$ ($W$: unrenormalized band width)
for the onset of the Stoner instability as a function of the band
filling $n$ for an $s$ band and: (a) constant density of states,
(b) sc lattice, (c) bcc lattice, and (d) fcc lattice. Different
lines correspond to: $V=J=0$ (dotted lines); $V=0.15U$, $J=0$
(dashed lines); $V=0.15U$, $J=0.03U$ ($J'=J$) (solid lines). }
\label{fig:sto}
\end{figure}

The study of the stability of the PM and FM states can be done
analytically for the constant density of states. For this density
of states the magnetic energy (Eq.(\ref{emagg})) is a quadratic
function of $m^2$:
\begin{equation}
E_{\rm mag}(m)=
\left[A(n)+B\left(\frac{m^2}{4}\right)\right]\left(\frac{m^2}{4}\right),
\label{emag}
\end{equation}
with, assuming $J=J'$,
\begin{eqnarray}
\label{emaga}
A(n)&=&W-U,                                            \nonumber \\
 &-&\frac{1}{2}z\left[V(3n^2\!-\!6n\!+\!2)-J(n^2\!-\!2n\!-\!4)\right],
                                                                 \\
\label{emagb}
B&=&-z(V-3J).
\end{eqnarray}
and $W=2zt$. For a given value of $n$, the minimum of $E_{mag}(m)$
depends on the actual parameters $A(n)$ and $B$. The PM state is
unstable against the FM state when $A(n)<0$; in the absence of
inter-site interactions the well known Stoner criterion for the
constant density of states, $U>W$ is recovered. In the general
case with $V\neq 0$ and $J\neq 0$ the value of $U/W$ above which
the PM state becomes unstable depends on $n$, as seen in Fig.
\ref{fig:sto}(a).

As shown above, the inter-site Coulomb interaction $V$ promotes FM
states for low numbers of electrons or holes, while around half
filling ($0.5\lesssim n \lesssim 1.5$), it tends to stabilize the
PM state. The inter-site exchange matrix element $J$ always favors
the FM state since $n^2-2n-4<0$ when $0<n<2$.

If $A(n)<0$ and $B>0$, as considered by Hirsch\cite{Hir99} who
assumes $J>V$, the PM state is unstable, and $E_{mag}(m)$ has a
minimum at $m_m$. If $m_m<m_s$ the most stable solution is a
non-saturated FM state. In this case the condition
$dE_{mag}(m)/dm=0$, with $E_{mag}$ given in Eq. (\ref{emag}), is
equivalent to the condition of equal Fermi energies for the
majority and minority spin sub-bands considered instead in Ref.
\onlinecite{Hir99}. If $m_m>m_s$, the most stable solution is the
SF state.

However, for realistic V and J parameters\cite{Str94,Cormier}
one has
$V>3J$, thus $B$ is negative. In this condition if $A(n)>0$ $E_{mag}(m)$
has a minimum at $m=0$, a maximum at $m_M$ and vanishes at $m=m_0$.
When $m_0>m_s$ the PM state is stable and when $m_0<m_s$ the SF state
is the most stable solution but the PM state is metastable. If
$A(n)<0$ the PM state becomes unstable while the SF state remains
the most stable solution. Thus if $B<0$ weak FM states are
rigorously excluded and, as $U/W$ increases, the SF state is
stable before the Stoner instability of the PM state. The critical
value of $U/W$ above which the SF state becomes stable is given by:
\begin{equation}
A(n)+\frac{1}{4}Bm_s^2=0.
\end{equation}

\begin{figure}
\includegraphics*[width=9cm]{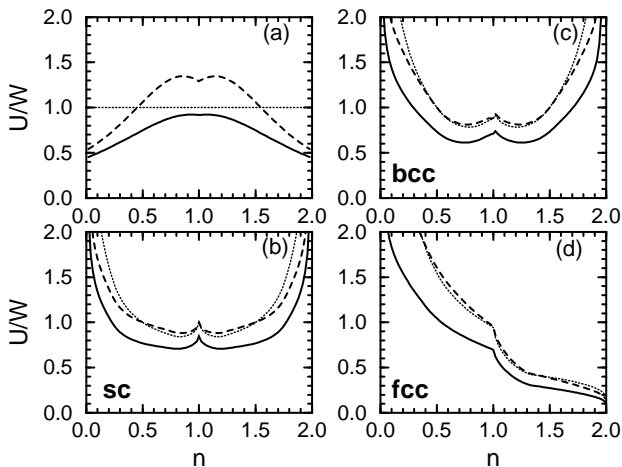}
\caption {Critical value of $U/W$ ($W$: unrenormalized band width)
above which the SF phase becomes stable as a function of the band
filling $n$ and for: (a) constant density of states, (b) sc lattice,
(c) bcc lattice, and (d) fcc lattice. The meaning of different
lines and the parameters are the same as in Fig.
\protect\ref{fig:sto}. }
\label{fig:fer}
\end{figure}

The variation with $n$ of this critical value is shown in Fig.
\ref{fig:fer} for the same set of parameters as in Fig. \ref{fig:sto}(a),
and it can be verified that these curves are always {\it below\/} the
corresponding ones for the Stoner instability, except at both ends
where the values of $U/W$ are the same.

\subsection{Magnetic instabilities in cubic lattices}

In the previous section we have discussed the FM instability for an $s$
band assuming a constant density of states. However this density of
states cannot be associated with any existing lattice. We now consider
the case of cubic lattices: simple , body centered and face centered
cubic lattices with a hopping integral $-\beta$ limited to nearest
neighbors. The corresponding dispersion relation is then:
\begin{equation}
E({\bf k})= -\beta \sum_j \exp(i{\bf k.R}_j),
\end{equation}
where ${\bf R}_j$ ($j=1,...z)$ denote the set of vectors connecting an
atom to its nearest neighbors. In this case the density of states, and
consequently $I(n_{\sigma})$, must be calculated numerically. The
densities of states have been obtained from this dispersion relation by
carrying out the summation over the Brillouin zone using the linear
tetrahedron method.\cite{tetra}

However, in order to account accurately for the singularities of
$N(E)$ we have also used the analytical expressions given by
Jelitto \cite{Jelitto69} which approximate the actual densities of
states with an excellent relative error (less than $10^{-4}$). The
function $I(n_{\sigma})$ is derived from Eq. (\ref{Isigma}) and
its derivatives $I'(n_{\sigma})$ and $I''(n_{\sigma})$, which are
necessary to study the Stoner instability (Eq. (\ref{stoner})) are
determined by means of the relations (\ref{IPIPP}). In the
following we will always assume $J=J'$ and $V>3J$.

\begin{figure}
\includegraphics*[width=7cm]{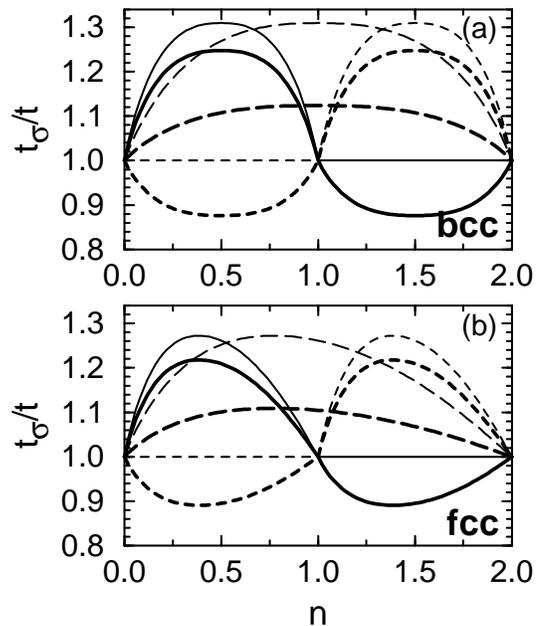}
\caption{Effective hopping integrals $t_{\sigma}$ (in units of $t$) as
a function of the band filling $n$ of an $s$ band for: (a) bcc and (b)
fcc lattices. Different lines refer to: PM phase (long dashed lines),
$t_{\uparrow}/t$ (full lines) and $t_{\downarrow}/t$ (dashed lines) in
the SF phase (stable at sufficiently large $U$, see
Fig. \protect\ref{fig:fer}). The parameters are: $V=0.15W$ ($W=16t$),
$J'=J$, and: $J=0$ (thin lines), $J=0.03W$ (heavy lines). }
\label{fig:teff}
\end{figure}

The effective hopping integral in the PM state $t_{\rm PM}$ has a
maximum at $n=1$ for the sc and bcc lattices, for which $N(E)$ has
a particle-hole symmetry, while the maximum is shifted to $n\simeq
0.76$ in the case of fcc lattice having asymmetric $N(E)$ [see
Fig. \ref{fig:teff}]. In all cases, the effective hopping integral
is reduced by $J>0$. At different filling of spin sub-bands, as
for instance in SF states, the effective up- and down-spin hopping
elements and the corresponding bandwidths are different. For
instance, for $n<1$ the up-spin bandwidth increases when this
sub-band is gradually filled, has a maximum at $n\simeq 0.38$ for
the fcc lattice, and then decreases back to the unrenormalized
value at $n=1$, while the down-spin sub-band is unrenormalized
(narrowed) when $J=0$ ($J>0$), as shown in Fig. \ref{fig:teff}(b).
The renormalization of spin sub-bands is interchanged for $n>1$
when the up-spin sub-band is filled and thus weakly narrowed in
the SF states.

To illustrate the general trends we have determined the Stoner
instabilities for sc, bcc, and fcc lattices using Eq. (\ref{stoner}),
and for three sets of parameters: (i) $V=J=\!0$, (ii) $V=0.15U$, $J=0$,
and (iii) $V=0.15U$, $J=0.03U$. The results are given in Fig.
\ref{fig:sto} and are in perfect agreement with the qualitative
predictions derived above (see Sec. IIB). Indeed, $V$ tends to stabilize
the FM state for low and high band fillings, while the reverse is found
around half-filling. Moreover, when $J$ is taken into account, it acts
in favor of FM states for any band filling.
This result confirm the earlier findings of
Hirsch.\cite{Hir89,Hir99}

\begin{figure}
\includegraphics*[width=8.5cm]{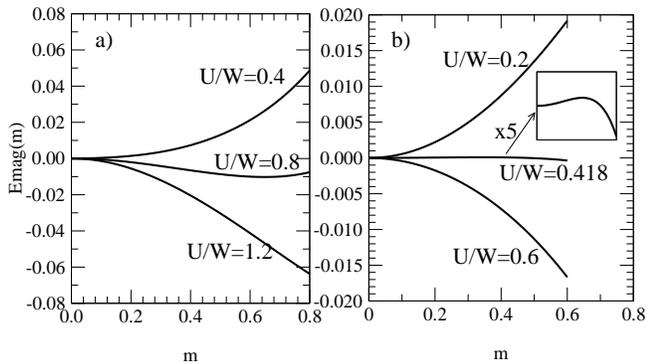}
\caption{Magnetic energy $E_{mag}$ (in units of the unrenormalized
band width W) as a function of the magnetic moment $m$ for a fcc
lattice with an $s$ band for two band fillings (a)$n=0.9$ (b)
$n=1.4$ and increasing values of $U/W$. The other parameters are
$V=0.15U$, $J=J'=0$. The inset shows the case $n=1.4$, $U/W=0.418$
with an enlarged energy scale, proving that the PM state is
metastable.} \label{fig:emag_m}
\end{figure}

The regions of stability of the SF phase are shown in Fig.
\ref{fig:fer}. When $n$ approaches 0 or 2 the curves of Figs.
\ref{fig:sto} and \ref{fig:fer} corresponding to the same values of $V$
and $J$ become closer and closer to each other since the magnetic moment
is infinitesimal in both limits and the second order expansion of
$I_{\sigma}$ is valid for any value of $m\leq m_s$. Furthermore, it is
found that for all lattices and band fillings, except for the fcc
lattice with $n\gtrsim 1.3$, the Stoner instability is found for a
smaller value of $U/W$ than that needed to stabilize the SF state. This
means that, contrary to the case of the constant density of states, the
PM state is never metastable except for the fcc lattice with
$n\gtrsim 1.3$ where a very narrow domain of metastability exists. This
is illustrated for the fcc lattice in Fig. \ref{fig:emag_m} where we
have plotted $E_{mag}(m)$ for several values of $U/W$ ($V=0.15U$,
$J=J'=0$) and two values of the band filling: $n=0.9$ and $1.4$. It is
clearly seen that, when $n=0.9$, as $U/W$ increases the most stable
phase is successively the PM, unsaturated FM and SF phase while, when
$n=1.4$, the PM phase is immediately followed by the SF phase but
there is a narrow range of $U/W$ where the PM phase is metastable.

\begin{figure}
\includegraphics*[width=7cm]{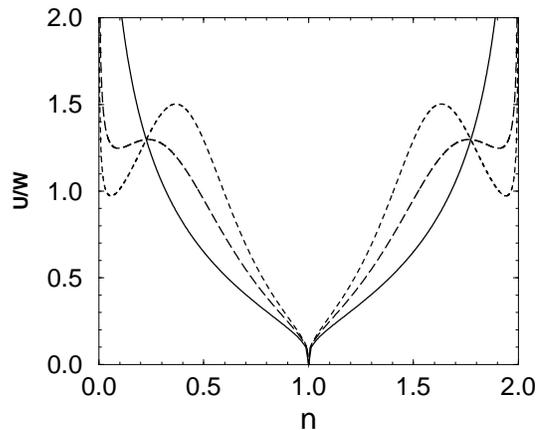}
\caption {Critical value of $U/W$ ($W$: unrenormalized band width)
for the onset of the Stoner instability in a bcc lattice with an
$s$ band as a function of the band filling $n$ for increasing
values of $V/U$ (0: full line, 0.35: dashed line, 0.55: dotted line) and
$J=J'=0$.}
\label{fig:sto_V}
\end{figure}

It is important to realize that the above results were derived
using a rather small value of $V/U$ and, consequently, the effect of $V$
is also small. However, it increases rapidly with $V$ as shown in Fig.
\ref{fig:sto_V} for the Stoner instability in the fcc lattice. Note that
the onset of the Stoner instability is independent of $V$ for two band
fillings since, as seen in Eq. (\ref{eq:emag2}), $E_{mag}$ does not
depend on $V$ when ${I'}_0^2+I_0{I''}_0=0$.

Finally, we must emphasize that all our results discussed so far
have been obtained assuming interactions limited to first nearest
neighbors. When hopping integrals between farther neighbors are taken
into account, the study of the influence of interatomic Coulomb
integrals on the electronic structure becomes more involved. Indeed,
whereas in the first nearest neighbor case the $I_{\sigma}$ function in
Eq. (\ref{ts}) can be calculated once for all since the density of
states $N(E)$ scales with the hopping integral without changing its
shape, this is no longer true when farther neighbor interactions are
included and the solution of the problem
becomes more tricky. Thus we have carried it out only in the study of
the realistic valence $spd$ band of transition metals presented in the
next section.

\section{ Ferromagnetism for hybridized $spd$ bands }

In the previous section we presented the effect of interatomic Coulomb
interactions on the electronic structure and their influence on the
onset of ferromagnetism in a tight-binding $s$ band with interactions
limited to first nearest neighbors. From this study we can draw several
conclusions. First, the relative numerical values of the parameters are
critical to determine the FM instabilities. Then the shape of the
density of states has also a strong influence. Thus in order to derive
reliable conclusions for FM transition metals we must now
generalize this model to hybridized $s$, $p$ and $d$ bands with farther
interactions as well as realistic Coulomb matrix elements.

\subsection{ The Hamiltonian }

In the basis of real $s, p$, and $d$ atomic orbitals (denoted by
$\lambda$ and $\mu$ indices) the Hamiltonian is determined by the
bare atomic levels $\varepsilon_{\lambda}$, the bare hopping
integrals $-t_{i\lambda,j\mu}$, and by the matrix elements of the
Coulomb interaction. In the following we keep the most important
of these latter terms which can be selected by comparing the
numerical values of all the Coulomb matrix elements obtained using
explicit expressions of the $s, p$, and $d$ atomic orbitals. From
these calculations,\cite{Cormier} it turns out that the leading
intra-atomic Coulomb matrix elements are:
\begin{eqnarray}
\label{uspd} U_{\lambda\mu} &=& \langle\phi_{i\lambda}({\bf r})
\phi_{i\mu}({\bf r'}) | \frac{1}{|{\bf r-r'}|} |
             \phi_{i\lambda}({\bf r}) \phi_{i\mu}({\bf r'}) \rangle ,   \\
\label{jspd} J_{\lambda\mu} &=& \langle\phi_{i\lambda}({\bf r})
\phi_{i\mu}({\bf r'}) |\frac{1}{|{\bf r-r'}|} |
             \phi_{i\mu}({\bf r}) \phi_{i\lambda}({\bf r'}) \rangle,
\end{eqnarray}
i.e., the intra-atomic Coulomb and exchange integrals involving
two orbitals that we already introduced in a previous paper.
\cite{Barreteau00} These parameters satisfy an important relation
for any pair of orbitals with the same orbital quantum number $l$:
\begin{equation}
U_{\lambda \lambda} = U_{\lambda\mu}+2J_{\lambda\mu}.
\label{eq:ull}
\end{equation}

Note also that all $U_{\lambda\lambda}$ elements are equal for
orbitals $\lambda$ with the same $l$. In what follows we consider
six Coulomb integrals: $U_{ss}$, $U_{sp}$, $U_{sd}$, $U_{pp'}$,
$U_{pd}$, $U_{dd'}$, and five exchange integrals: $J_{sp}$,
$J_{sd}$, $J_{pp'}$, $J_{pd}$, $J_{dd'}$, with $p'\ne p$ and
$d'\ne d$, i.e., we have taken the average value of some sets of
Coulomb integrals; for instance, $U_{dd'}$ is the average of all
Coulomb integrals involving two different $d$ orbitals. The
corresponding values of $U_{pp}$ and $U_{dd}$ are determined from
Eq. (\ref{eq:ull}).

The most important Coulomb interatomic interactions are:
\begin{equation}
\label{vspd} V^{\lambda\mu}_{ij} = \langle\phi_{i\lambda}({\bf r})
\phi_{j\mu}({\bf r'}) | \frac{1}{|{\bf r-r'}|} |
             \phi_{i\lambda}({\bf r}) \phi_{j\mu}({\bf r'}) \rangle,
\end{equation}
i.e.,the electrostatic inter-site interactions. From atomic
orbital calculations,\cite{Cormier} it is found that their bare
value (before screening) is almost independent of the considered
pair of orbitals and close to $e^2/R_{ij}$, $R_{ij}$ being the
spacing between atoms $i$ and $j$, so that we can approximate them
by:
\begin{equation}
V_{ij}=  V_0 \frac{R_0}{R_{ij}} \quad , \quad \forall R < R_c,
\end{equation}
where $R_0$ is a reference distance which is chosen to be the first
nearest neighbor bulk equilibrium spacing, $V_0$ is a parameter to be
determined in order to take screening effects into account, and $R_c$ a
cut-off distance. We must stress that in this calculation the other
two-site matrix elements involving four $d$ orbitals (which are the
most important in this problem), and in particular the exchange
integrals, are at least two orders of magnitude smaller than $V_{ij}$.

The $spd$ band Hamiltonian is then written as follows,
\begin{eqnarray}
\label{hspd} H &=&-\sum_{i\lambda,j\mu,\sigma \atop i\ne j}
      t_{i\lambda,j\mu}a_{i\lambda\sigma}^{\dagger}a_{j\mu\sigma}
  +\sum_{i\lambda\sigma}\varepsilon_{\lambda}n_{i\lambda\sigma}
 \nonumber \\
&+& \sum_{i \lambda  } U_{\lambda \lambda}
 n_{i \lambda\uparrow} n_{i \lambda\downarrow} +
 \frac{1}{2} \sum_{i \lambda \mu ,\lambda \ne \mu \atop \sigma \sigma' }
 U_{\lambda \mu} n_{i \lambda \sigma}n_{i \mu \sigma'}  \nonumber    \\
&+& \frac{1}{2} \sum_{i \lambda \mu,\lambda \ne \mu \atop \sigma
\sigma'} J_{\lambda \mu}  a^{\dagger}_{i \lambda
\sigma}a^{\dagger}_{i \mu \sigma'}
          a^{}_{i \lambda \sigma'}a^{}_{i \mu \sigma}     \nonumber   \\
&+& \frac{1}{2} \sum_{i\lambda\mu,\lambda\ne\mu \atop\sigma }
 J_{\lambda\mu}a^{\dagger}_{i\lambda\sigma}a^{\dagger}_{i\lambda-\sigma}
               a^{}_{i\mu-\sigma}a_{i\mu\sigma}     \nonumber     \\
 &+& \frac{1}{2} \sum_{ij\lambda\mu,i\ne j \atop\sigma\sigma' }
 V_{ij}a^{\dagger}_{i \lambda
\sigma}a^{\dagger}_{j \mu \sigma'}
          a_{j\mu \sigma'}a^{}_{i \lambda \sigma}.
\end{eqnarray}
$a_{i\lambda\sigma}^{\dagger}$ and $a_{i\lambda\sigma}$ are the
creation and annihilation operators of an electron in the
spin-orbital $|i\lambda\sigma\rangle$; $n_{i\lambda\sigma}$ is the
corresponding occupation number operator. The above multi-band
Hubbard Hamiltonian can be solved in the framework of the HFA (Eq.
(\ref{hfa})) which leads to a one particle TBHF Hamiltonian:
\begin{eqnarray}
H_{\rm HF} &=& \sum_{i\lambda\sigma} \varepsilon_{i\lambda\sigma}
                              n_{i\lambda\sigma}
+ \sum_{i\lambda\mu,\lambda\ne\mu \atop\sigma} h_{i\lambda, i\mu, \sigma}
 a^{\dagger}_{i \lambda \sigma} a_{i \mu \sigma}     \nonumber   \\
&-& \sum_{i \lambda j \mu \sigma \atop i\ne j} t_{i\lambda,j\mu\sigma}
 a^{\dagger}_{i \lambda \sigma} a_{j \mu\sigma}-E_{dc},
\end{eqnarray}
which is the direct generalization of Eq. (\ref{hfs})
except for the appearance of new terms $h_{i\lambda, i\mu,
\sigma}$, i.e., on-site but interorbital hopping integrals. The
latter terms vanish in the bulk for cubic symmetry, and also in
hexagonal symmetry except for small $sp$ contributions. They
should be taken into account when the symmetry is
reduced.\cite{Piv96} The renormalized matrix elements are given
by:
\begin{eqnarray}
\label{erenorm}\varepsilon_{i\lambda\sigma} &=&
\varepsilon_{\lambda}+ U_{\lambda \lambda} \langle n_{i
\lambda-\sigma}\rangle + \sum_{\mu\ne\lambda
\atop\sigma'}U_{\lambda\mu}
\langle n_{i\mu\sigma'}\rangle                         \nonumber \\
&-&\sum_{\mu\ne\lambda}
 J_{\lambda\mu} \langle n_{i \mu \sigma}\rangle
+\sum_j V_{ij}\langle n_j\rangle, \\
h_{i\lambda,i\mu\sigma} &=& -U_{\lambda\mu}\langle a^{\dagger}_{i\mu\sigma}
a_{i\lambda\sigma}\rangle  \nonumber \\
&+& J_{\lambda\mu}(\langle a^{\dagger}_{i\mu\sigma}
a_{i\lambda\sigma}\rangle+2\langle a^{\dagger}_{i\mu-\sigma}
a_{i\lambda-\sigma}\rangle),  \\
\label{trenorm}t_{i\lambda,j\mu \sigma} &=& t_{i\lambda,j\mu} +
V_{ij}\langle a^{\dagger}_{j\mu\sigma}a_{i\lambda\sigma}\rangle.
\end{eqnarray}

Note that there is a renormalization of on-site levels
due to the charge interaction $V_{ij}$, but in an homogeneous
system where each site has the same charge $\langle n_j\rangle=n$,
it simply produces a rigid shift of the levels and can therefore
be ignored for bulk calculations. Actually the most important
effect of $V_{ij}$ is the renormalization of hopping integrals, and
thus of the bandwidth, which is different for up and down spins in
the FM case.

\subsection{ The parametrization of the TBHF Hamiltonian }

In order
to perform realistic calculations for the FM $3d$ transition
metals, we combined the tight-binding approach of Mehl and
Papaconstantopoulos\cite{Pap96} for the PM state with corrections
originating from the electron-electron interactions in the FM
states.
At this point, it is important to note that for any bulk
geometrical configuration the HF renormalization of the energy
levels and of the hopping integrals in the PM state are implicitly
included in the parameters of Mehl and Papaconstantopoulos.
\cite{Pap96} Indeed, these parameters have been fitted on
electronic structure calculations carried out in the density
functional formalism. It is therefore convenient to take the PM
state as a reference. The Hamiltonian H can then be written in the
following way:
\begin{equation}
H=H_0 + \Delta H - \Delta E_{dc},
\end{equation}
where $H_0$ is the PM Hamiltonian parametrized in
Ref. \onlinecite{Pap96} and $\Delta H$ is the perturbation due to
the onset of ferromagnetism. More details can be found in
Ref. \onlinecite{Barreteau00}. Finally, $\Delta E_{dc}$ is the
variation of double counting terms between the FM and PM states.
Note that this parametrizatiom assumes a non-orthogonal basis set.
Consequently $n=\langle n_j\rangle$ is the net population at site
$j$ and not the band filling which is given by the gross atomic
population as defined in Ref. \onlinecite{Christoffersen89}.
However, it is this latter population which should be used to
calculate the magnetic moment per atom.

The Coulomb ($U_{\lambda\mu}$) and exchange ($J_{\lambda\mu}$)
integrals have been determined from their atomic values and then
reduced by appropriate screening factors as explained in Ref.
\onlinecite{Barreteau00}. It is at present not possible to get a
well controlled procedure which would describe the screening of
the atomic interactions when atoms build a crystal. Therefore, we
introduce two multiplicative screening factors, $\alpha_U$ and
$\alpha_J$, operating on the $U$ and $J$ atomic values,
respectively. It is known that Coulomb interactions are strongly
screened, while exchange interactions remain almost
unscreened.\cite{vdM88} This is the reason why we have kept the
same value of $\alpha_J$ ($\alpha_J=0.70$) as in Ref.
\onlinecite{Barreteau00}. Let us now discuss the values of
$\alpha_U$ and $V_0$. From Sec. II it is clear that the
interaction $V_{ij}$ modifies the onset of ferromagnetism and,
consequently, $\alpha_U$ and $V_0$ should be determined in a
correlated way. In addition these parameters should not vary
significantly between Fe, Co and Ni which have almost the same
interatomic spacing. We will see in Sec. IV that the values
$\alpha_U=0.12$ and $V_0=0.5eV$ lead to bulk spin magnetic moments
close to the experimental values and to bulk electronic structures
(in particular bandwidths and splitting of the two spin sub-bands)
in good agreement with local spin density calculations for the
three ferromagnetic 3d elements. Note that the resulting values of
$U_{\lambda\mu}$ and $J_{\lambda\mu}$ are small (see Table I) as
usual in the HFA to simulate the correlation effects and that the
ratio $U_{dd}/V_0\simeq 4$ seems quite reasonable.

\begin{table}
\caption{
Coulomb $U_{\lambda \mu}$ and exchange $J_{\lambda \mu}$
on-site integrals (in eV) obtained from an atomic calculation and
screened, respectively, by the factors: $\alpha_U=0.12$ and
                                        $\alpha_J=0.7$. The
intraorbital integrals are given by:
$U_{pp}=U_{pp'}+2J_{pp'}$, and $U_{dd}=U_{dd'}+2J_{dd'}$.}
\vskip .7cm
\begin{ruledtabular}
\begin{tabular}{ccccc||cccc}
$U_{\lambda\mu}$ & Fe & Co & Ni & & $J_{\lambda\mu}$ & Fe & Co &
Ni \cr \colrule $U_{ss}$   & 0.263 & 0.284 & 0.304 & &   -       &
- &  -    &  -    \cr $U_{sp}$   & 0.158 & 0.170 & 0.182 & &
$J_{sp}$  & 0.184 & 0.198 & 0.213 \cr $U_{sd}$   & 0.367 & 0.367 &
0.417 & & $J_{sd}$  & 0.105 & 0.104 & 0.101 \cr $U_{pp'}$  & 0.158
& 0.170 & 0.182 & & $J_{pp'}$ & 0.230 & 0.248 & 0.266 \cr $U_{pd}$
& 0.294 & 0.294 & 0.334 & & $J_{pd}$  & 0.084 & 0.084 & 0.081 \cr
$U_{dd'}$  & 0.823 & 0.886 & 0.950 & & $J_{dd'}$ & 0.571 & 0.595 &
0.625 \cr
\end{tabular}
\end{ruledtabular}
\label{tab:uj}
\end{table}

\section{Application to ferromagnetic transition metals}
\label{sec:3d}

We have performed TBHF calculations on the three FM $3d$ transition
metals Fe(bcc), Co(hcp) and Ni(fcc) at their experimental equilibrium
structure, i.e, $a_{bcc}=2.87$ \AA for Fe(bcc), $a_{hcp}=2.51$ \AA,
$c/a_{hcp}=1.62$ for Co(hcp), and $a_{fcc}=3.52$ \AA for Ni(fcc).
The cut-off radius $R_c$ for the interatomic Coulomb interaction was
chosen between second and third nearest neighbors.

\subsection{Ferromagnetic states of bcc Fe }
\label{sec:fe}

\begin{figure}
\includegraphics*[width=7cm]{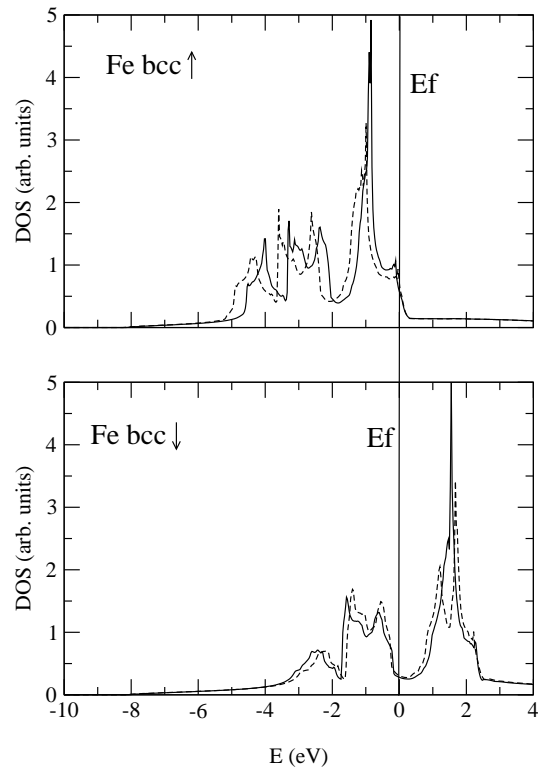}
\caption {Density of states as obtained for bcc Fe in the
tight-binding Hartree-Fock model with (solid lines, $V_0=0.5$eV)
and without (dashed lines) Coulomb inter-site interaction, for
$\uparrow$-spin electrons (top), and $\downarrow$-spin electrons
(bottom) in the FM ground state.} \label{fig:DOS_Fe_V_noV}
\end{figure}

As we have shown in Sec. II, the main effect of the Coulomb
inter-site interaction is to modify the width of the majority spin
band with respect to the minority one. To illustrate this effect
in the case of Fe we have performed a self-consistent TBHF
calculation with and without this interaction, as shown in Fig.
\ref{fig:DOS_Fe_V_noV}. It appears very clearly that the bandwidth
of the majority spin $d$ electrons is significantly smaller than
that of the minority one when the Coulomb interaction is "switched
on".

\begin{figure}
\includegraphics*[width=7cm]{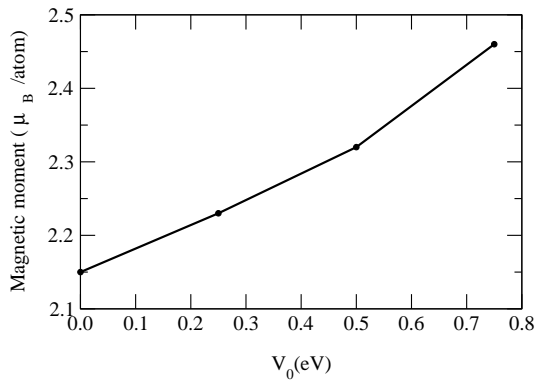}
\caption {Variation of the magnetic moment of bcc Fe as a function
of the Coulomb inter-site interaction $V_0$.} \label{fig:mag_Fe_V}
\end{figure}

The other effect of the Coulomb inter-site interaction is to
modify the Stoner instability. In particular, it was shown in Sec.
II in the analytic treatment of the $s$ band model that $V_{ij}$
tends to play in favor of the destabilization of the PM states for
nearly filled bands. Consequently we expect an increase of the
magnetic moment when $V_{ij}$ increases. This can be seen in Fig.
\ref{fig:mag_Fe_V}, where the evolution of the magnetic moment is
plotted for different values of $V_0$ ranging from 0 to 0.75 eV
and fixed values of $U_{\lambda\mu}$ corresponding to
$\alpha_U=0.12$. The spin magnetic moment obtained for $V_0=0.5eV$
is in very good agreement with the experimental results
\cite{Kittel} (see Table II).

\begin{figure}
\includegraphics*[width=7cm]{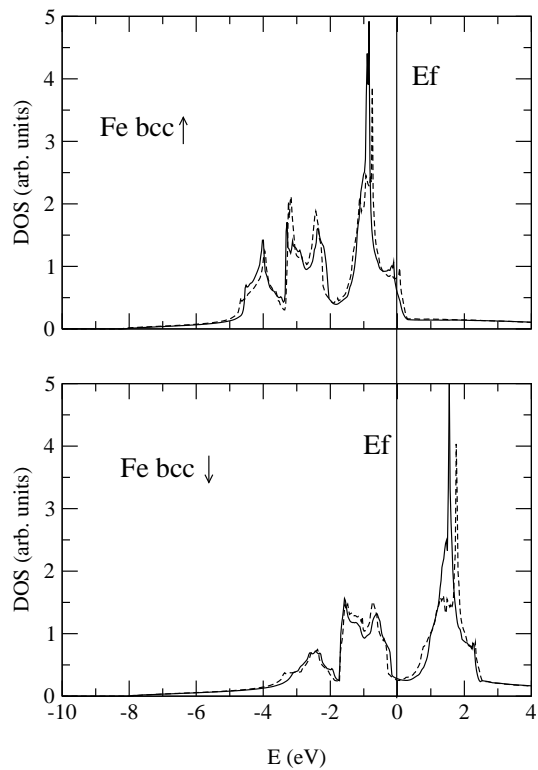}
\caption {Density of states as obtained for bcc Fe in the
tight-binding Hartree-Fock model (solid lines, $V_0=0.5$ eV), and
in the band structure calculation using the FLAPW method of
Ref. \onlinecite{Wien} (dashed lines) for $\uparrow$-spin electrons
(top), and $\downarrow$-spin electrons (bottom) in the FM ground
state.} \label{fig:DOS_Fe_TB_WIEN}
\end{figure}

\begin{figure}
\includegraphics*[width=9cm]{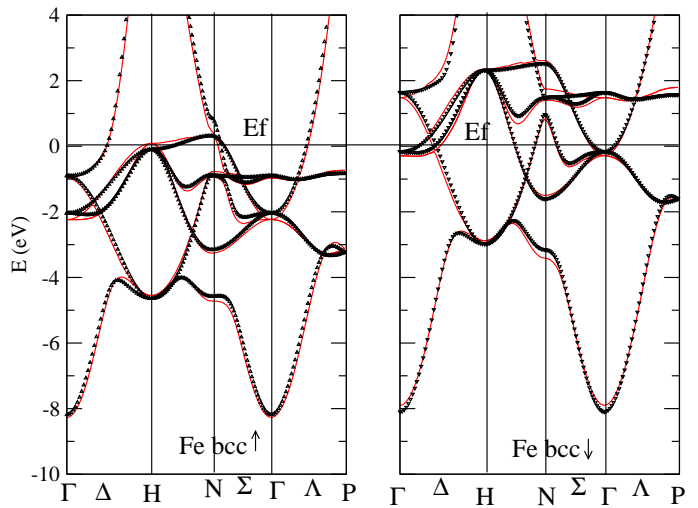}
\caption {Electronic structure as obtained for bcc Fe in the
tight-binding Hartree-Fock model ($V_0=0.5$eV, black triangles),
and in the band structure calculation using the FLAPW method of
Ref. \onlinecite{Wien} (solid lines) for $\uparrow$-spin electrons
(left), and $\downarrow$-spin electrons (right) in the FM ground
state.} \label{fig:band_Fe_TB_WIEN}
\end{figure}

Finally, in Figs. \ref{fig:DOS_Fe_TB_WIEN} and
\ref{fig:band_Fe_TB_WIEN} we have compared the densities of states
and band structures as obtained from our TBHF calculation with
$V_0=0.5eV$ and from an FLAPW LSDA calculation using the WIEN
code.\cite{Wien} The agreement is almost perfect proving that the
set of intra and inter-site Coulomb and exchange interactions that
we chose, not only reproduces integrated quantities such as the
magnetic moment (see Table II) but is also able to describe very
accurately the splitting and change of bandwidth between majority
and minority spins (see Table III).

\subsection{Ferromagnetic states of Co}
\label{sec:co}

\begin{figure}
\includegraphics*[width=7cm]{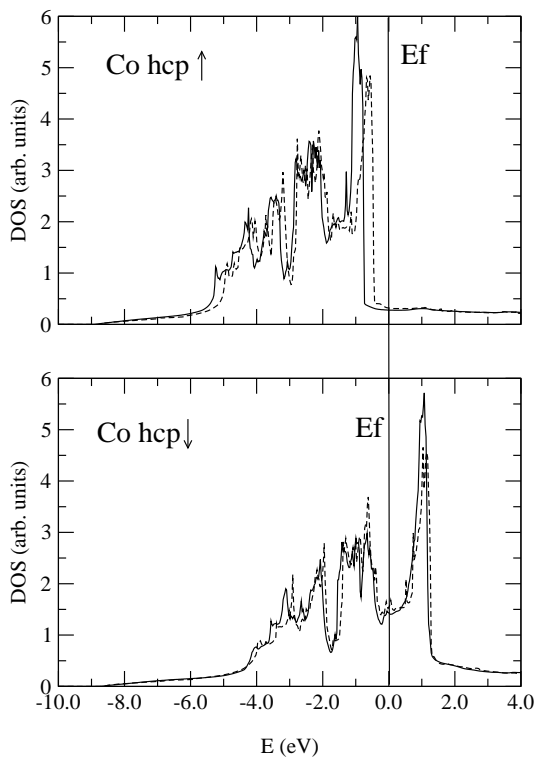}
\caption
{Same caption as in Fig. \protect\ref{fig:DOS_Fe_TB_WIEN} but for hcp
Co.}
\label{fig:DOS_Co_hcp}
\end{figure}

\begin{figure}
\includegraphics*[width=7cm]{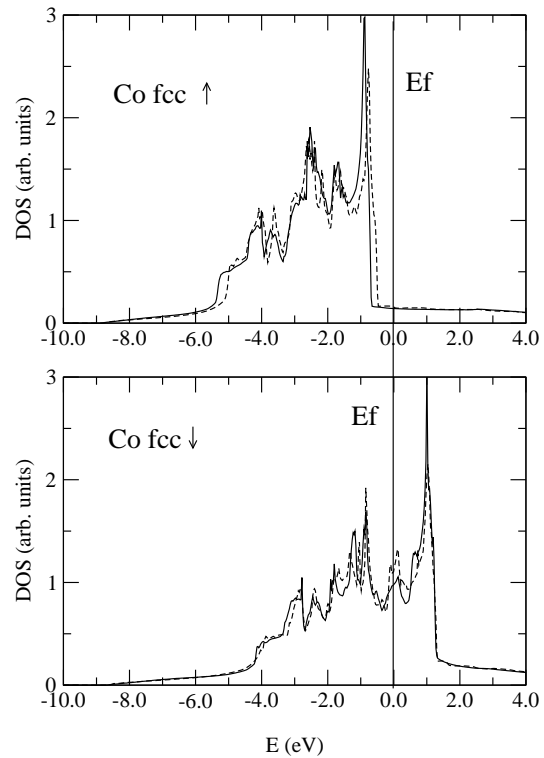}
\caption
{Same caption as in Fig. \protect\ref{fig:DOS_Fe_TB_WIEN}, but for fcc
Co.}
\label{fig:DOS_Co_fcc}
\end{figure}

We now present our results for Co (hcp and fcc) keeping the same
values for $\alpha_U$ and $V_0$ as for Fe. The results of our TBHF
and FLAPW LSDA calculations on hcp Co are shown in Fig.
\ref{fig:DOS_Co_hcp}, where the densities of states obtained with
the two methods are represented. Once again the agreement is
excellent for the magnetic moment (see Table II) and for the shape
and the width of the majority and minority spin densities of
states. In particular the difference in bandwidths between
majority spin and minority spin $d$ electrons is found to be
almost the same as with the WIEN code (see Table III). However,
there is a small quasi rigid shift of the $d$ band for the majority
spin density of states. Note that, Co being a saturated ferromagnet,
this small shift has almost no influence both on the magnetic moment
and on the total energy. We have also carried out TBHF and FLAPW LSDA
calculations on fcc Co with a lattice parameter $a_{fcc}=3.55$ \AA,
the densities of states are presented in Fig. \ref{fig:DOS_Co_fcc},
showing the same type of agreement between the two methods.

\subsection{Ferromagnetic states of fcc Ni }
\label{sec:ni}

\begin{figure}
\includegraphics*[width=7cm]{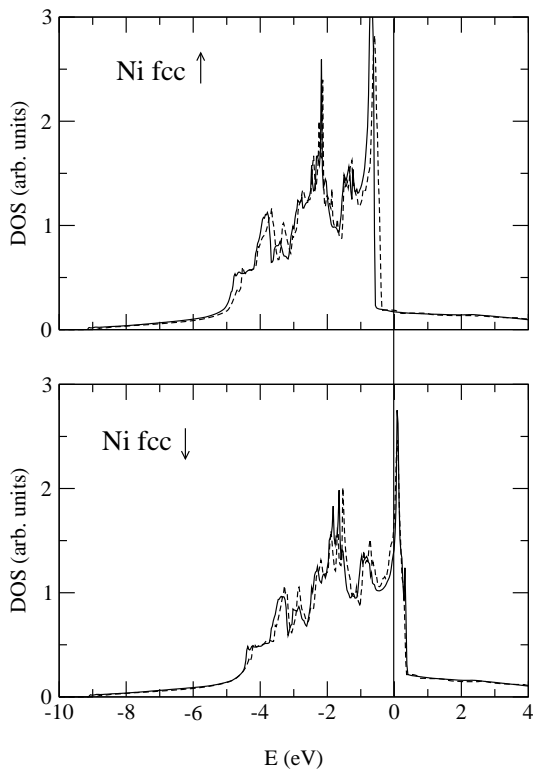}
\caption
{Same caption as in Fig. \protect\ref{fig:DOS_Fe_TB_WIEN}, but for fcc
Ni.}
\label{fig:DOS_Ni_fcc}
\end{figure}

The same values of $\alpha_U$ and $V_0$ were also used for Ni. The
result is extremely convincing since the magnetic moment is
exactly the same with TBHF and the WIEN code (see Table II), and
Fig. \ref{fig:DOS_Ni_fcc} shows an excellent agreement for the
electronic structure. As for cobalt, there is a slight shift of
the majority spin $d$ sub-band without consequences on the
magnetic moment and total energy since Ni is also a saturated
ferromagnet, but the shape of the densities of states and the
changes of the bandwidths (Table III) are very similar.

\begin{table}
\caption{Comparison of the spin magnetic moments (in Bohr magnetons)
obtained from tight-binding Hartree-Fock (TBHF) method, and the WIEN
code, compared with experimental values
\protect\cite{Kittel} for Fe, Co and Ni.}
\vskip .7cm
\begin{ruledtabular}
\begin{tabular}{cccc}
Element & TBHF & WIEN  & exp.  \\
\colrule
 Fe      & 2.32 & 2.23 & 2.13  \\
 Co(hcp) & 1.60 & 1.51 & 1.57  \\
 Co(fcc) & 1.59 & 1.60 &  --   \\
 Ni      & 0.58 & 0.58 & 0.56  \\
\end{tabular}
\end{ruledtabular}
\end{table}

\begin{table}
\caption{ Relative difference in $d$ bandwidths $(W_{d
\downarrow}-W_{d \uparrow})/\langle W_{d}\rangle$, in percentage,
as obtained in the tight-binding Hartree-Fock (TBHF) and WIEN code
calculations. $\langle W_d\rangle$ is the average bandwidth of
both spins.} \vskip .7cm
\begin{ruledtabular}
\begin{tabular}{ccc}
Element & TBHF & WIEN \\
\colrule
 Fe     & 15  & 12  \\
 Co(hcp)& 16  & 17  \\
 Ni     & 11  & 7  \\
\end{tabular}
\end{ruledtabular}
\label{tab:band_width}
\end{table}

To conclude this section, the introduction of the inter-site
Coulomb interaction and the subsequent renormalization of the
hopping integrals in the $spd$ TBHF model has enabled us to obtain
an excellent overall agreement with calculations based on the
density functional formalism for the band structure, the density
of states and the magnetic moment of the three $3d$ FM elements.
However the splitting between up and down spin bands is
systematically slightly larger than in local spin density
calculations. This difference could be expected since with TBHF
the self-interaction is forbidden, as it should, while it is
allowed in the WIEN code, as usual in the density functional
theory. Indeed, if the self-interaction term is included in Eq.
(\ref{erenorm}), i.e., $\langle n_{i\lambda-\sigma}\rangle$ is
replaced by $\langle n_{i\lambda-\sigma}\rangle+\langle
n_{i\lambda\sigma}\rangle$, the term proportional to
$U_{\lambda\lambda}$ no more contributes to the splitting between
up and down spin bands.

\section{Conclusions}

To summarize, we have used a tight-binding Hartree-Fock model
including the renormalization of the hopping integrals due to
inter-site Coulomb interactions in order to put forward its
influence on the appearance of ferromagnetism. First, we
reconsidered the model of non-degenerate $s$ band and found a
generalized Stoner criterion (Eqs. (17-18)). As we have shown, the
renormalization of the hopping integrals which originates from the
inter-site Coulomb elements strongly modify the conditions for
ferromagnetism. In agreement with earlier
studies,\cite{Hir89,Hir90,Hir96,Hir99} ferromagnetism is favored
for nearly filled or empty bands by the nearest neighbor Coulomb
interactions. As the actual FM instabilities are rather sensitive
to the system parameters, an accurate description of the density
of states and realistic interaction parameters are of crucial
importance to understand the behavior of $3d$ transition metals.

Next we have shown that the behavior found for the non-degenerate $s$
band model has important consequences in realistic transition metals.
We extended the model to the case of hybridized $s, p$ and $d$ bands
and used it
to investigate the electronic structure of FM Fe, Co and Ni. It was
found that the width of the majority spin band is
always smaller than that of the minority spin one,
as obtained in electronic structure calculations performed by
{\it ab initio} methods.
An excellent
overall agreement (band structure, densities of states, magnetic
moment) with the local spin density calculations is obtained for
the three elements.

Finally, it has to be emphasized that this renormalization of the
hopping integrals is also present in the non-magnetic case and is
a function of the environment of the pair of atoms involved in the
hopping. Here we have only considered the bulk geometry. It would
be interesting to study the case of surfaces and especially of
small clusters in which the effect of the change of environment is
expected to be the strongest.

\acknowledgments
We are very grateful to D. Cormier and C. Minot who calculated all
intra- and intersite matrix elements of the Coulomb interaction
in Fe$_2$ and Ni$_2$ helping us to select the most important ones.
It is also our pleasure to thank F. Flores, J. E. Hirsch, G. Hug,
O. Jepsen, G. A. Sawatzky, and G. Stollhoff for valuable discussions.
A.M.O. acknowledges the kind hospitality of DSM/DRECAM/SPCSI, Centre
d'Etudes de Saclay, where part of this work was completed, and the
support by the Polish State Committee of Scientific Research (KBN),
Project No. 5~P03B~055~20.


\end{document}